\documentclass[twocolumn]{aastex701}
\usepackage{amsmath,amssymb}
\usepackage{graphicx}% Include figure files
\usepackage{dcolumn}% Align table columns on decimal point
\usepackage{bm}% bold math
\usepackage{xcolor}
\usepackage{multirow}

\begin{document}

\title{Constraining the Nanohertz Gravitational Wave Background with an X-ray Pulsar Timing Array from \textit{NICER} observations}

\correspondingauthor{Shi-jie Zheng}
\email[show]{zhengsj@ihep.ac.cn}

\correspondingauthor{Shu-Xu Yi}
\email[show]{sxyi@ihep.ac.cn}

\author[orcid=0009-0000-1101-3470,gname=Tian-Yong, sname=Cao]{Tian-Yong Cao} 
\affiliation{State Key Laboratory of Particle Astrophysics, Institute of High Energy Physics, Chinese Academy of Sciences, Beijing 100049, China}
\affiliation{University of Chinese Academy of Sciences, Chinese Academy of Sciences, Beijing 100049, People’s Republic of China}
\email{}

\author[orcid=0000-0003-2256-6286,gname=Shijie, sname=Zheng,gname=Shi-Jie, sname=Zheng]{Shi-jie Zheng} 
\affiliation{State Key Laboratory of Particle Astrophysics, Institute of High Energy Physics, Chinese Academy of Sciences, Beijing 100049, China}
\affiliation{University of Chinese Academy of Sciences, Chinese Academy of Sciences, Beijing 100049, People’s Republic of China}
\email{}

\author[orcid=0000-0001-7599-0174, gname=Shu-Xu, sname=Yi]{Shu-Xu Yi} 
\affiliation{State Key Laboratory of Particle Astrophysics, Institute of High Energy Physics, Chinese Academy of Sciences, Beijing 100049, China}
\affiliation{University of Chinese Academy of Sciences, Chinese Academy of Sciences, Beijing 100049, People’s Republic of China}
\email{}

\author[orcid=0000-0002-2749-6638, gname=Ming-Yu, sname=Ge]{Ming-Yu Ge} 
\affiliation{State Key Laboratory of Particle Astrophysics, Institute of High Energy Physics, Chinese Academy of Sciences, Beijing 100049, China}
\affiliation{University of Chinese Academy of Sciences, Chinese Academy of Sciences, Beijing 100049, People’s Republic of China}
\email{}

\author[gname=Yi-Tao, sname=Yin]{Yi-Tao Yin} 
\affiliation{State Key Laboratory of Particle Astrophysics, Institute of High Energy Physics, Chinese Academy of Sciences, Beijing 100049, China}
\affiliation{University of Chinese Academy of Sciences, Chinese Academy of Sciences, Beijing 100049, People’s Republic of China}
\email{}

\author[gname=Yao-Ming, sname=Duan]{Yao-Ming Duan} 
\affiliation{State Key Laboratory of Particle Astrophysics, Institute of High Energy Physics, Chinese Academy of Sciences, Beijing 100049, China}
\affiliation{University of Chinese Academy of Sciences, Chinese Academy of Sciences, Beijing 100049, People’s Republic of China}
\email{}

\author[gname=Xiang Yang, sname=Wen]{Xiang Yang, Wen} 
\affiliation{State Key Laboratory of Particle Astrophysics, Institute of High Energy Physics, Chinese Academy of Sciences, Beijing 100049, China}
\affiliation{University of Chinese Academy of Sciences, Chinese Academy of Sciences, Beijing 100049, People’s Republic of China}
\email{}

% \author[orcid=0000-0000-0000-0001,sname='North America']{Tundra North America}
% \altaffiliation{Kitt Peak National Observatory}
% \affiliation{University of Saskatchewan}
% \email[show]{fakeemail1@google.com}  

% \collaboration{all}{The Terra Mater collaboration}

\begin{abstract}

\noindent We present constraints on the nanohertz gravitational wave background (GWB) using X-ray pulsar timing data from the Neutron Star Interior Composition Explorer(\textit{NICER}). By analyzing six millisecond pulsars over a six-year observational baseline, we employed a Bayesian framework to model noise components and search for a common red signal consistent with a GWB from supermassive black hole binaries (assuming a spectral index $\gamma_{\rm gwb}=13/3$). Our results show no significant evidence for a GWB, yielding a 95\% upper limit of $\log_{10}(A_{\rm gwb})<-13.4$. Weak evidence for Hellings-Downs spatial correlations was found (S=2.5), though the signal remains statistically inconclusive. Compared to radio and $\gamma$-ray pulsar timing arrays, the \textit{NICER} constraint is currently less stringent but 
demonstrates the feasibility of X-ray timing with \textit{NICER} for GWB studies and highlights the potential for improved sensitivity with future X-ray missions.

\end{abstract}

\keywords{\uat{Gravitational waves}{678} ---
          \uat{Pulsars}{1306} ---
          \uat{X-ray astronomy}{1810} ---
          \uat{Bayesian statistics}{1900} ---
          \uat{Neutron stars}{1108} ---
          \uat{Astrophysical black holes}{98}}


\section{Introduction}

With the advent of gravitational-wave astronomy, it has become possible to explore the cosmic population of massive binaries across the Universe. As galaxies evolve and merge over cosmic time \citep{white1978core}, the supermassive black holes (SMBHs) residing at their centers are expected to form binary systems \citep{begelman1980massive}, which emit GWs as they inspiral. Due to the vast number of galaxies in the Universe, the GWs from these SMBH binaries superpose incoherently, giving rise to a stochastic gravitational wave background (GWB) \citep{rajagopal1994ultra,wyithe2003low,ravi2015prospects,burke2019astrophysics}. This GWB encodes valuable information about the history of galaxy mergers and the dynamics of binary SMBHs \citep{jaffe2003gravitational,sesana2008stochastic,sesana2013insights}, making it one of the most compelling GW sources to study. Moreover, additional nHz GWs are predicted from exotic sources such as cosmic strings \citep{kibble1976topology,damour2000gravitational,siemens2007gravitational,olmez2010gravitational,sanidas2012constraints}, cosmological phase transitions \citep{caprini2010detection,xue2021constraining}, and inflation in the early Universe \citep{grishchuk2005relic,zhao2013constraints,lasky2016gravitational,galtier2017turbulence}, offering powerful probes of fundamental physics and cosmology.

Pulsar Timing Arrays (PTAs) are currently the only mature method available to detect GWs in the nHz frequency band \citep{xu2023searching, gold1969rotating, jenet2005detecting, 2025RAA....25c5022C, 2014MNRAS.445.1245Y, 2016SCPMA..5989511Y, 2010arXiv1008.1782C, 1979ApJ...234.1100D, 1983ApJ...265L..39H, 2003ApJ...583..616J, 2006ApJ...653.1571J, 2004ApJ...606..799J, 2005ApJ...625L.123J, 2011MNRAS.414.3251L, 2001ApJ...562..297L, 1978SvA....22...36S, 2008MNRAS.390..192S, 2009MNRAS.394.2255S, 2011ApJ...730...29W, 2010MNRAS.407..669Y}. Owing to the remarkable rotational stability of pulsars, their beams sweep across the Earth with predictable regularity, allowing precise measurements of pulse times of arrival (ToAs). When a GW passes through the light path between the pulsars and the earth, it perturbs spacetime, causing subtle shifts in the ToAs. Unlike white and red noise, which can arise from statistical fluctuations, instrumental effects, interstellar medium variations, intrinsic pulsar spin irregularities, etc., a GWB induces a characteristic spatial correlation in the timing residuals between different pulsars. This angular correlation, first derived by \cite{hellings1983upper}, is known as the Hellings-Downs (HD) curve, and serves as a key signature for distinguishing a GWB from uncorrelated noise.

\begin{figure*}
    \centering
    \includegraphics[width=\linewidth]{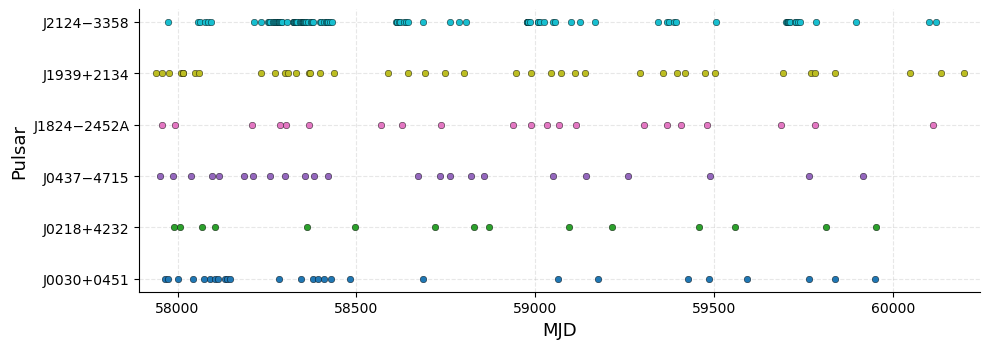}
    \caption{\textbf{Observation epochs of six NICER millisecond pulsars}: Each horizontal line represents the observation span of one pulsar, with markers indicating the epochs of available ToA measurements.}
    \label{mjds}
\end{figure*}

In 2023, the world’s major radio PTA collaborations jointly released their latest constraints on the GWB \citep{agazie2023nanograv,antoniadis2023second,reardon2023search,xu2023searching}. The Parkes Pulsar Timing Array (PPTA) constrained the GWB amplitude to $2.04_{-0.22}^{+0.25} \times 10^{-15}$ \citep{reardon2023search}, while the Chinese Pulsar Timing Array (CPTA) provided the strongest evidence to date for HD spatial correlations, reaching a statistical significance of 4.6$\sigma$ \citep{xu2023searching}. Additionally, the \textit{Fermi-LAT} collaboration, using a $\gamma$-ray-based PTA, placed an upper limit on the GWB amplitude of $6.7\times10^{-15}$ \citep{kerr2024upgrading}.

The Neutron Star Interior Composition Explorer (\textit{NICER}) is a dedicated X-ray observatory designed to study emissions from neutron stars (NSs), with primary goals of constraining the NS mass-radius relation and investigating their high-energy emission mechanisms. Leveraging its exceptional timing precision, \textit{NICER} has enabled high-accuracy measurements of X-ray pulsars, thereby providing valuable constraints on the amplitude of the GWB from PTA in the X-ray band.

In Section \ref{Data and Methods}, we will describe the processing of \textit{NICER} data and the methods used to constrain the GWB. The constraint results and their statistical significance will be presented in Section \ref{Results}. Finally, we will conclude with a summary and outlook in Section \ref{Conclusion and Outlook}.

\section{Data and Methods}
\label{Data and Methods}
\subsection{\textit{NICER} data reduction and timing analysis}

\textit{NICER} have monitored many X-ray millisecond pulsars (MSPs) with more than six years of observations. The six MSPs, including PSRs J1939+2134, J1824$-$2452A, J0437$-$4715, J0030+0451, J0218+4232, and J2124$-$3358, demonstrate high X-ray fluxes, small rotation periods, and minimal period derivatives, coupled with high stability (although PSR J1939+2134 and B1821$-$24 show low-frequency timing irregularities dominate the timing residuals \citep{1994ApJ...428..713K,2009MNRAS.400..951V}), rendering them excellent candidates for GWB detection. The exposures of PSRs J1939+2134, J1824$-$2452A, J0437$-$4715, J0030+0451, J0218+4232 and J2124$-$3358 are around several mega-seconds over the period from June 2017 to September 2023 (see Table 1 in \cite{2024Univ...10..174Z} and Figure \ref{mjds}) , which could be utilized to supply more accurate timing results. 

For each pulsar, the observational data is filtered firstly with the selection criteria outlined in \cite{2024Univ...10..174Z}, and then the arrival time of each photon at the local observatory is corrected to the Solar System Barycenter (SSB) with the pulsar's ephemeris from the International Pulsar Timing Array (IPTA) Data Release 2 (DR2) \citep{2019MNRAS.490.4666P}. Therefore, we yields the high-precision "standard" pulse profile using six years of data. Then the entire data are segmented based on photon count. Each segment comprises approximately 50,000 to 200,000 photons within a 30-day span. We obtain pulsed profiles for each segment and calculate ToAs through cross-correlation analysis with the standard profile. The ToA errors is calculated by performing Gaussian sampling across the pulse profile. The pulsar period parameters and timing residuals are then refined using Tempo2 \citep{2006MNRAS.369..655H, 2006MNRAS.372.1549E}. 

\subsection{Noise Modeling}
Similar to \(\gamma\)-ray pulsars \citep{kerr2024upgrading, fermi2022gamma}, X-ray pulsars emit high-energy photons, making them largely unaffected by plasma dispersion effects during propagation. While dispersion delays, characterized by the dispersion measure (DM), significantly limit the timing precision of low-energy radio pulsars \citep{cordes2016frequency, donner2020dispersion}, the propagation speed of X-ray and higher-energy photons is essentially frequency-independent. Consequently, the timing noise of X-ray pulsars can be broadly classified into two categories: white noise and red noise.

\begin{table}
\centering
\renewcommand{\arraystretch}{1.5}
\caption{\textbf{Table of White Noise Parameters for Different Pulsars}: The white noise parameters are directly obtained from \texttt{PINT} fitting. For pulsars where the inclusion of the EQUAD parameter is not appropriate, or where the fitted EQUAD value is negligibly small, we set its logarithmic value to $-10$.
}
\setlength{\tabcolsep}{20pt}
\begin{tabular}{ccc}
\hline
            & EFAC & $\log_{10}$EQUAD \\ \hline
J0030+0451  & 0.89 & -6.00      \\
J0218+4232  & 1.10 & -10        \\
J0437-4715  & 0.91 & -5.98      \\
J1824-2452A & 4.50 & -10        \\
J1939+2134  & 1.40 & -10        \\
J2124-3358  & 1.21 & -5.87      \\ \hline
\end{tabular}
\label{white_noise}
\end{table}

White noise is typically dominated by instrumental statistical noise and uncertainties in the measurement of pulse ToAs, characterized by temporally uncorrelated random fluctuations. For each observed ToA measurement uncertainty $\sigma_{\rm ToA}$, we apply a scaling factor to account for possible misestimation of its amplitude, and introduce an additional white noise term to model excess instrumental noise or other unknown white noise contributions \citep{edwards2006tempo2}:
\begin{equation}
    C_{{\rm white}, {\rm I}}(t_i, t_j) = \delta_{ij}\big({\rm EFAC}_{\rm I}^2\sigma_i^2+{\rm EQUAD}_{\rm I}^2\big)\,,
\end{equation}
where $i,j$ denote the indices of the ToAs, I represents the index of the pulsar, $\sigma_i$ is the measurement uncertainty of the $i$-th ToA, and $\delta_{ij}$ is the Kronecker delta.

The power spectral density of red noise is defined as:
\begin{equation}
    P_{\rm I}(f) = {A_{\rm I}^2 \over 12\pi^2}({f \over f_c})^{-\gamma_{\rm I}}\,\rm yr^{-3}\,.
\end{equation}
Here, $A_{\rm I}$ denotes the red noise amplitude for each pulsar, and $\gamma_{\rm I}$ is the spectral index. $f_c$ is the reference frequency, which we set to $\rm yr^{-1}$ in this work. Existing observational data suggest that the spectral index typically falls within the range of 2 to 7 \citep{alam2020nanograv, goncharov2021identifying}. The covariance matrix of the red noise is defined as:
\begin{equation}
    C_{{\rm red}, {\rm I}}(t_i, t_j) = \int_0^\infty P_{\rm I}(f)\cos{[2\pi f(t_i-t_j)]}\mathrm{d}f\,.
\end{equation}

Finally, we obtain the total noise covariance matrix by summing the red noise and white noise covariance matrices for each pulsar and concatenating the covariance matrices of all pulsars:
\begin{equation}
    C_n = 
    \begin{pmatrix}
    C_{{\rm white}, {\rm I}} & 0 & \cdots \\
    0 & C_{{\rm white}, {\rm II}} & \cdots \\
    \vdots & \vdots & \ddots \\
    \end{pmatrix}
    + \begin{pmatrix}
    C_{{\rm red}, {\rm I}} & 0 & \cdots \\
    0 & C_{{\rm red}, {\rm II}} & \cdots \\
    \vdots & \vdots & \ddots \\
    \end{pmatrix}\,.
\end{equation}
The dimension of this matrix is equal to the total number of ToAs across all pulsars, denoted by $N_{\rm ToA}$.

\begin{table}
\centering
\renewcommand{\arraystretch}{1.5}
\caption{\textbf{Table of Priors for MCMC Fitting Parameters}: References: [1]~\citet{antoniadis2023secondII}; 
[2]~\citet{caballero2016noise}; 
[3]~\citet{goncharov2021identifying}; 
[4]~\citet{hazboun2022detection}.}
\setlength{\tabcolsep}{5pt}
\begin{tabular}{cccc}
\hline
pulsar                       & parameter    & prior                & reference         \\ \hline
\multirow{2}{*}{J0030+0451}  & $\log_{10}A$ & Normal(-14.9, 1.1)  & \multirow{2}{*}{[1],[2]} \\
                             & $\gamma$     & Normal(5.49, 1.93)  &                   \\
\multirow{2}{*}{J0218+4232}  & $\log_{10}A$ & Normal(-14.1, 1.7)  & \multirow{2}{*}{[2]} \\
                             & $\gamma$     & Normal(3.90, 1.70)  &                   \\
\multirow{2}{*}{J0437-4715}  & $\log_{10}A$ & Normal(-14.4, 0.1)  & \multirow{2}{*}{[3]} \\
                             & $\gamma$     & Normal(2.02, 0.30)  &                   \\
\multirow{2}{*}{J1824-2452A} & $\log_{10}A$ & Normal(-12.6, 0.5)  & \multirow{2}{*}{[4]} \\
                             & $\gamma$     & Normal(4.11, 1.83)  &                   \\
\multirow{2}{*}{J1939+2134}  & $\log_{10}A$ & Normal(-13.9, 0.1)  & \multirow{2}{*}{[3]} \\
                             & $\gamma$     & Normal(1.53, 0.42)  &                   \\
\multirow{2}{*}{J2124-3358}  & $\log_{10}A$ & Uniform(-18,-8)  & \multirow{2}{*}{$\diagup$} \\
                             & $\gamma$     & Uniform(0,7)  &                   \\
\multirow{2}{*}{GWB}         & $\log_{10}A$ & Uniform(-18, -10)  & \multirow{2}{*}{$\diagup$} \\
                             & $\gamma$     & Constant(13/3)  &  \\ \hline
\end{tabular}
\label{prior}
\end{table}

\subsection{Gravitational Wave Background Modeling}
We assume that the GWB is generated by the superposition of a large number of SMBHB. For each individual source, the amplitude, frequency, and phase are random and stochastic; however, their collective contribution results in a common power spectral density \citep{sesana2004low}:
\begin{equation}
    P_{\rm gwb}(f) = {A_{\rm gwb}^2 \over 12\pi^2}({f \over f_c})^{-\gamma_{\rm gwb}}\,\rm yr^{-3}\,.
\end{equation}
Here, $A_{\rm gwb}$ denotes the amplitude of the GWB, and $\gamma_{\rm gwb}$ is its spectral index. For GWBs originating from this source population, the spectral index has a theoretical value of $\gamma_{\rm gwb} = 13/3$ \citep{sesana2004low, ligo2017basic}. Similarly, this power spectral density can be converted into a common covariance matrix $\tilde{C}_{{\rm gwb},\rm IJ}(t_{{\rm I}i}, t_{{\rm J}j})$ shared by all pulsars, with a dimension of $N_{\rm ToA}$.

When a GW passes through the light path between the pulsars and the earth, they simultaneously affect the timing signals of different pulsars. However, since the GWB arises from an isotropic superposition of sources with random polarization and phase, contributions from different directions tend to cancel out on average, leaving only a residual correlation that depends on the angular separation between pulsar pairs. This correlation is described by the HD curve \citep{hellings1983upper}:
\begin{equation}
    \Gamma_{\rm IJ}(\theta_{\rm IJ}) = ({3\over2}x\ln{x}-{1\over4}x+{1\over2})\delta_{\rm IJ}\,,
\end{equation}
where $x=\frac{1-\cos{\theta_{\rm IJ}}}{2}$, $\theta_{\rm IJ}$ denotes the angular separation between pulsars I and J. 

After incorporating the HD curve, the total covariance matrix induced by the GWB is given by:
\begin{equation}
    C_{\rm gwb} = \Gamma_{\rm IJ}\tilde{C}_{{\rm gwb},\rm IJ}\,.
\end{equation}

\subsection{Sampling Method}

In this work, we use \texttt{Enterprise} \citep{enterprise, enterprise_extensions} to compute the likelihood and \texttt{Eryn} \citep{Karnesis:2023ras, michael_katz_2023_7705496, 2013PASP..125..306F} to perform MCMC sampling. The likelihood is defined as:
\begin{equation}
    \mathcal{L} = \frac{1}{\sqrt{(2\pi)^{N_{\rm ToA}}}|C_p|}\exp{(-{1\over2}\delta t_p^TC_p^{-1}\delta t_p)}\,,
    \label{likelihood}
\end{equation}
where $C_p = C_n + C_{\rm gwb}$ represents the total covariance matrix, as defined in the previous subsections. $\delta t_p$ denotes the vector obtained by concatenating the timing residuals of all pulsars.

With the above model, the parameters associated with $\mathcal{L}$ include the timing model parameters, the white noise parameters and the red noise parameters for each pulsar, and the common red noise (CRN) parameters representing the GWB. With high-quality radio pulsar timing data \citep{agazie2023nanograv, antoniadis2023second, reardon2023search, xu2023searching}, it is also possible to constrain the spectral index observationally. However, for $\gamma$-ray and X-ray pulsars, current data are insufficient to establish the presence of such a background signal. Therefore, in this work, we fix $\gamma_{\rm gwb} = 13/3$ and derive an upper limit on the amplitude $A_{\rm gwb}$.

Since the GWB signal follows a power-law spectrum similar to intrinsic red noise, their parameter spaces are expected to be correlated. Therefore, we also allow the red noise parameters for each pulsar, $A_I$ and $\gamma_{\rm I}$, to vary freely in the fit. In contrast, white noise and timing model parameters have a negligible impact on the GWB parameters. We therefore fix these parameters to their best-fit values obtained using \texttt{TEMPO2} \citep{hobbs2006tempo2, edwards2006tempo2, hobbs2009tempo2} and \texttt{PINT} \citep{luo2021pint, susobhanan2024pint}, and treat them as constants in our analysis. The white noise parameters for each pulsar are listed in the Table \ref{white_noise}.

The red noise properties of individual pulsars have been extensively studied in previous works. For some pulsars, observational data are available not only in the X-ray band but also in the $\gamma$-ray or radio bands. We incorporate the constraints on red noise from these studies as priors in our analysis. The priors for all fitted parameters are summarized in the Table \ref{prior}.

\begin{figure}
    \centering
    \includegraphics[width=\linewidth]{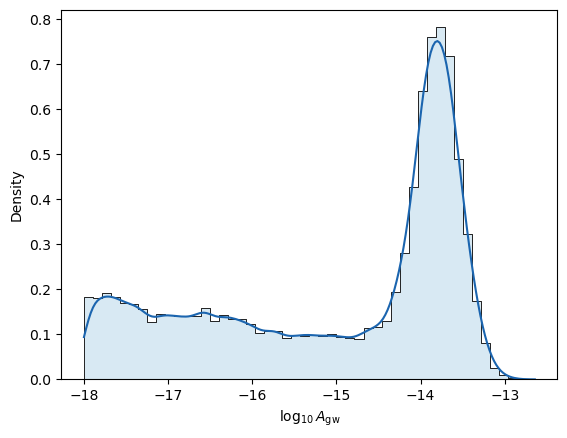}
    \caption{\textbf{Log-posterior distribution of $A_{\rm gwb}$ derived from \textit{NICER} data}.}
    \label{log10_Agw}
\end{figure}

\section{Results}
\label{Results}

Using \textit{NICER} data, with the likelihood defined by Equation \ref{likelihood} and the priors given in Table \ref{prior}, we obtain the posterior distribution of the GWB amplitude $A_{\rm gwb}$ through MCMC sampling, as shown in Figure \ref{log10_Agw}. Although the logarithmic posterior distribution exhibits a peak around $-14$, we do not interpret this peak as evidence for the presence of a GWB signal. By computing the Bayes factor between models with and without a GWB component, we find that the model including a GWB is favored by only $\Delta \log Z = 1.47$, which does not constitute significant evidence. Furthermore, when the spectral index $\gamma_{\rm gwb}$ is allowed to vary freely, the posterior distribution fails to converge near the theoretically expected value of $\gamma_{\rm gwb} = 13/3$. Overall, the fitted amplitude primarily reflects the level of CRN at $\gamma_{\rm gwb} = 13/3$, suggesting that any potential GWB contribution should not exceed this level. Therefore, we adopt the 95\% upper bound of the posterior distribution of the red noise amplitude as the 95\% upper limit on the GWB amplitude from \textit{NICER} data, yielding $\log_{10}{A_{\rm gwb}} < -13.4$.

\begin{figure}
    \centering
    \includegraphics[width=\linewidth]{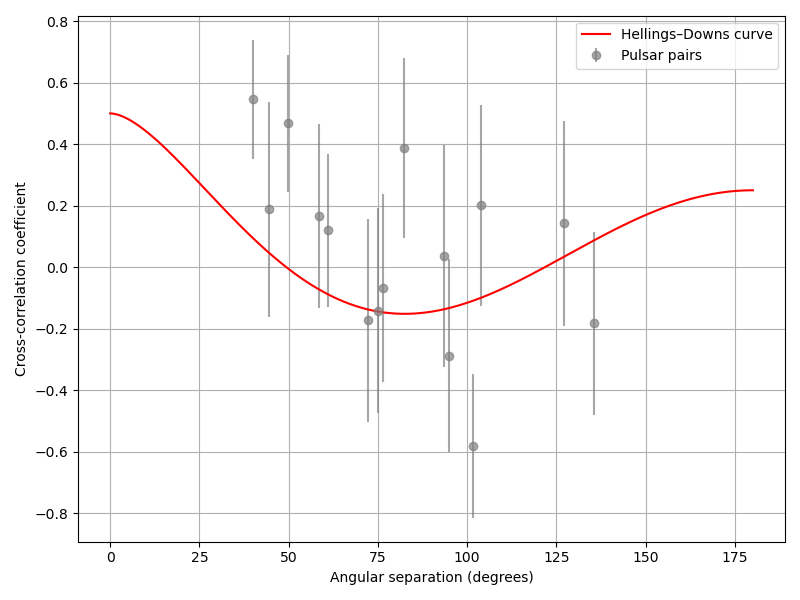}
    \caption{\textbf{Plot of the cross-correlation coefficients between different pulsar pairs as a function of their angular separation}.}
    \label{pulsar_correlations}
\end{figure}

To investigate the spatial correlation of the CRN, we plot the cross-correlation coefficients between different pulsar pairs as a function of their angular separation, as shown in Figure \ref{pulsar_correlations}. We use the frequentist method proposed by \cite{jenet2004constraining} to assess its statistical significance:
\begin{equation}
    \mathcal{S} = \sqrt{\frac{N(N-1)}{2}}\frac{\sum_{
m {\rm I}<
m J}(c_{\rm IJ}-\bar{c})(h_{\rm IJ}-\bar{h})}{\sqrt{\sum_{
m {\rm I}<
m J}(c_{\rm IJ}-\bar{c})^2\sum_{
m {\rm I}<
m J}(h_{\rm IJ}-\bar{h})^2}}\,,
\end{equation}
where $c_{\rm IJ}$ is the cross-correlation coefficient between pulsars I and J, $h_{\rm IJ}$ is the corresponding theoretical value from the HD curve, \(\bar{s}\) and the \(\bar{h}\) is the average defined as \(\bar{s},\bar{h}=\frac{2}{N(N-1)}\sum_{
m {\rm I}<
m J}s_{\rm IJ},h_{\rm IJ}\). And $N$ is the total number of pulsar pairs. Applying this method yields a statistical significance $\mathcal{S} = 2.5$, suggesting a potential correlation signal, though not yet significant. Since $\mathcal{S}$ approximately follows a standard normal distribution, $\mathcal{S}=2.5$ corresponds to a $\sim98.8\%$ confidence level (below the conventional $3\sigma$ threshold for detection).

We also compare the Bayesian evidence for models with a GWB including the HD correlation versus a GWB without HD correlation, and find a Bayes factor of $\Delta \log Z = 0.369$. According to the conventional Jeffreys scale \citep{jeffreys1998theory}, $\Delta \log Z < 1$ indicates inconclusive evidence, while values of $1$–$3$, $3$–$5$, and $>5$ correspond to weak, moderate, and strong evidence, respectively. Thus, our result does not provide significant support for the HD correlation.

\begin{figure}
    \centering
    \includegraphics[width=\linewidth]{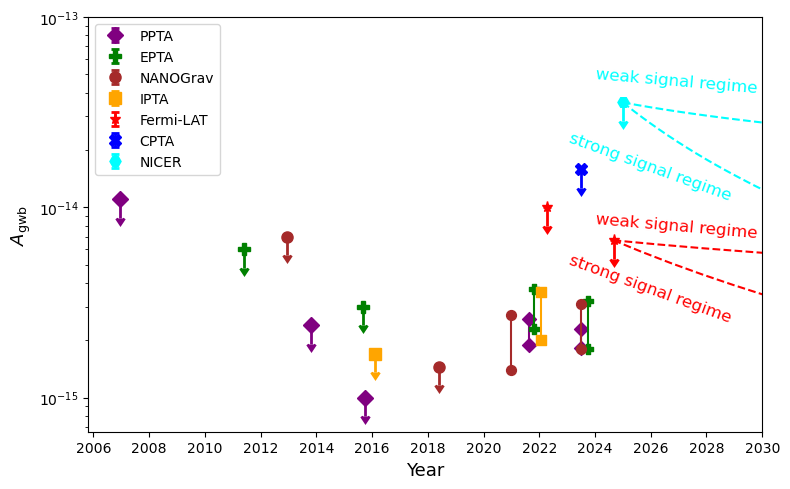}
    \caption{\textbf{Constraints on the GWB Amplitude from Different PTAs}: All GWB amplitude limits are shown at the reference frequency $f_c = 1\, \rm yr^{-1}$, assuming a spectral index of $\gamma_{\rm gwb} = 13/3$. The data are taken from the Parkes Pulsar Timing Array (PPTA) \citep{jenet2006upper, shannon2013gravitational, shannon2015gravitational, goncharov2021evidence, reardon2023search}, the European Pulsar Timing Array (EPTA) \citep{van2011placing, lentati2015european, chen2021common, antoniadis2023second}, the North American Nanohertz Observatory for Gravitational Waves (NANOGrav) \citep{demorest2012limits, arzoumanian2018nanograv, arzoumanian2020nanograv, agazie2023nanograv}, the International Pulsar Timing Array (IPTA) \citep{verbiest2016international, antoniadis2022international}, the Chinese Pulsar Timing Array (CPTA) \citep{xu2023searching}, and the \textit{Fermi-LAT} Pulsar Timing Array \citep{fermi2022gamma, kerr2024upgrading}. The time associated with each data point corresponds to the publication date of the respective paper. The cyan dashed line represents the projected improvement of GWB constraints from \textit{NICER} data over time.}
    \label{pta_total}
\end{figure}

After obtaining the upper limit on the GWB amplitude, we compare the constraint from \textit{NICER} data with those from PTAs in other wavebands, as shown in Figure \ref{pta_total} \footnote{The CPTA provides a credible interval for $\log A_{\rm gwb}=-14.4_{-2.8}^{+1.0}$ rather than just an upper limit. However, due to the large width of this interval, we only display the upper bound in the figure for clarity.} . Currently, the upper limit derived from \textit{NICER} is approximately one order of magnitude higher than the constraints from mainstream radio PTAs and about three times higher than the limit obtained from $\gamma$-ray pulsar timing. For an ideal PTA, the GWB amplitude limit is expected to improve following the relation $A_{\rm gwb} \propto T_{\rm obs}^{-\gamma_{\rm gwb}/2} = T_{\rm obs}^{-13/6}$ \citep{siemens2013stochastic,pol2021astrophysics}. We also show the projected improvement curve in Figure \ref{pta_total}.

In addition to the GWB produced by SMBHBs, other sources of the GWB with different spectral indices $\gamma_{\rm gwb}$ have been proposed. For example, relic GWs originating from scale-invariant inflation in the early Universe are expected to have $\gamma_{\rm gwb} = 5$ \citep{zhao2011constraint}, while the decay of cosmic strings is predicted to produce a power-law spectrum with $\gamma_{\rm gwb} = 16/3$ \citep{damour2005gravitational}.

We apply the same method to constrain the amplitudes of GWBs with different spectral indices $\gamma_{\rm gwb}$. We find that as the spectral index increases, the amplitude of the CRN tends to decrease. However, similar to the case with $\gamma_{\rm gwb} = 13/3$, if a GWB with the theoretically predicted spectral index exists, its amplitude remains below the level of the CRN observed in \textit{NICER} pulsars. The Bayes factors remain small and do not provide significant evidence for the presence of such a signal. Therefore, we adopt the 95\% upper bound of the posterior distribution as the 95\% confidence upper limit on the GWB amplitude for each $\gamma_{\rm gwb}$.

\begin{figure}
    \centering
    \includegraphics[width=\linewidth]{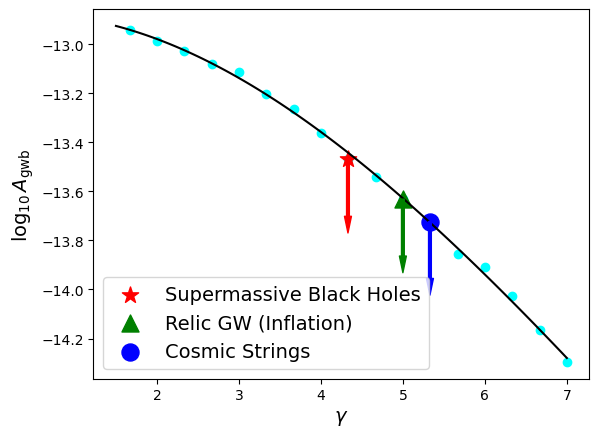}
    \caption{\textbf{95\% Confidence Upper Limits on GWB Amplitude for Different Spectral Indices}: The cyan points represent the 95\% confidence upper limits on the GWB amplitude obtained from MCMC fitting for different values of $\alpha$. The red pentagrams, green triangles, and blue circles correspond to GWBs from SMBHBs, scale-invariant inflation in the early Universe, and the decay of cosmic strings, respectively. The black curve shows the smoothed trend obtained using a cubic spline fit based on these data.}
    \label{A-gamma}
\end{figure}

The 95\% confidence upper limits on the amplitude $\log_{10}{A_{\rm gwb}}$ for different values of $\gamma_{\rm gwb}$ are shown in Figure~\ref{A-gamma}. The cyan points denote the upper limits obtained from MCMC fitting at each value of $\gamma_{\rm gwb}$, and the black curve represents the smoothed trend derived using a cubic spline interpolation. Therefore, even in scenarios without specific theoretical expectations for $\gamma_{\rm gwb}$, one can still infer the corresponding upper limits on the GWB amplitude across a broad range of spectral indices \citep{arzoumanian2021searching, khmelnitsky2014pulsar}.

\section{Conclusion and Outlook}
\label{Conclusion and Outlook}
In this work, we have presented a constraint on the GWB amplitude using X-ray pulsar timing data from \textit{NICER}, under the assumption of a fixed spectral index $\gamma_{\rm gwb} = 13/3$. Our Bayesian analysis yields a Bayes factor of $\Delta\log\mathcal{Z} = 1.47$, and the posterior distribution for the spectral index fails to converge near its theoretical value when allowed to vary. These findings suggest that the observed common-spectrum signal is more likely attributable to CRN processes intrinsic to the pulsars, rather than to a true GWB. Accordingly, we adopt a conservative 95\% upper limit of $\log_{10}A_{\rm gwb} < -13.4$, which remains approximately an order of magnitude weaker than those reported by current radio PTAs, and a factor of $\sim$4 weaker than the most stringent constraints from $\gamma$-ray PTAs.

In terms of spatial correlations, we compute the cross-correlation coefficients between pulsar pairs and compare them to the Hellings-Downs (HD) curve. The resulting significance of $S = 2.5$ provides a tantalizing hint of spatial coherence, though it falls short of the threshold required for a confident detection. The Bayesian comparison between models with and without HD correlations further supports this interpretation, yielding only a marginal improvement ($\Delta\log\mathcal{Z} = 0.369$).

Despite the limitations of current \textit{NICER} data, our results are complementary to radio and $\gamma$-ray PTAs, and represent the first constraint on the nHz GWB based on X-ray timing observations alone. The immunity of X-ray photons to dispersion and scattering in the interstellar medium simplifies the timing analysis and avoids frequency-dependent propagation effects that complicate radio timing. Furthermore, given the significant overlap between \textit{NICER} pulsars and those monitored by the \textit{Fermi-LAT} PTA, our analysis lays the groundwork for future joint multiwavelength studies, which can help disentangle CRN from genuine stochastic backgrounds and validate noise models employed in traditional PTA analyses. 

In addition, a natural extension of this approach would be to combine X-ray timing PTA data with radio-based PTA observations. As demonstrated by \cite{kerr2025future}, $\gamma$-ray pulsars not only exhibit strong potential for PTA studies, but also provide excellent synergy with radio pulsars owing to the largely uncorrelated noise properties between the two bands \citep{smith2023third}. It is therefore reasonable to expect that the inclusion of X-ray pulsars could further enhance this complementarity. While such a joint analysis would require careful treatment of the differing systematics and noise properties inherent to X-ray and radio timing measurements—such as dispersion measure variations, chromatic timing noise, and instrument-dependent calibration uncertainties—the largely independent noise budgets across these wavebands also provide a powerful cross-check. A consistent stochastic signal recovered across radio and X-ray PTAs would therefore offer strong evidence for a genuine GW origin, whereas discrepancies could help isolate band-specific noise processes. Consequently, a broad, multi-band PTA analysis incorporating radio, X-ray, and gamma-ray timing data holds significant promise for improving the robustness and statistical significance of future constraints on the stochastic GW background.

Looking ahead, the advent of next-generation X-ray observatories—such as the Advanced X-ray Imaging Satellite (AXIS) \citep{mushotzky2018axis}, the Advanced Telescope for High-Energy Astrophysics (Athena) \citep{nandra2013hot}, and the enhanced X-ray Timing and Polarimetry (eXTP) \citep{zhang2019enhanced,2025SCPMA..6819502Z}—will usher in a new era of precision timing in the high-energy domain. With enhanced sensitivity, broader sky coverage, and improved temporal resolution, these missions will not only enable long-term monitoring of existing millisecond pulsars but also increase the total number of suitable PTA sources. Based on the expected scaling $A_{\rm gwb} \propto T_{\rm obs}^{-13/6}$, we anticipate a substantial improvement in sensitivity over time. Moreover, the intrinsic advantage of X-ray timing in bypassing DM noise makes it especially promising for building a robust, independent PTA.

In conclusion, our study demonstrates the viability of using X-ray data from \textit{NICER} to constrain the GWB in the nanohertz regime, and paves the way for future multi-band PTA efforts that will play a vital role in unraveling the low-frequency GW universe.

\begin{acknowledgments}
\textbf{This work is supported by the National Key R\&D Program of China (2021YFA0718500) from the Minister of Science and Technology of China (MOST). The authors thank the support from the National Natural Science Foundation of China (grant Nos. 12373051 and 12333007) and the International Partnership Program of Chinese Academy of Sciences (grant No. 113111KYSB20190020). SXY acknowledges the support by the Institute of High Energy Physics (grant no. E32983U8). The authors also thank Heng Xu for helpful discussions and contributions to this work.}
\end{acknowledgments}

\begin{contribution}
%%This section gives authors the space to recognize author contributions. The text inside this environment is NOT counted towards the total word quanta. At a minimum, manuscripts are expected to include this text:

All authors contributed equally to the Terra Mater collaboration.

%% But authors are expected to provide more specific details, e.g. 
%%
%%SC was responsible for writing and submitting the manuscript.
%%WWM came up with the initial research concept and edited the manuscript.
%%OTS obtained the funding and edited the manuscript.
%%EBF provided the formal analysis and validation. He also edited the manuscript.
%%GEH Supervised the undergraduates, wrote the software and administers the project github and Zenodo repositories.
%%
%% Authors can use the Contributor Role Taxonomy (CRediT) at
%% for ideas on how write a good statement tailored to their needs.

\end{contribution}

\facilities{NICER}

\software{
    TEMPO2 \cite{hobbs2006tempo2, edwards2006tempo2, hobbs2009tempo2}
    PINT \cite{luo2021pint, susobhanan2024pint}
    Enterprise \cite{enterprise, enterprise_extensions}
    Eryn \cite{Karnesis:2023ras, michael_katz_2023_7705496, 2013PASP..125..306F}
    }


% \appendix

% \section{Appendix information}

%% For this sample we use BibTeX plus aasjournalv7.bst to generate the

\bibliography{sample701}{}

@ARTICLE{2025SCPMA..6819502Z,
       author = {{Zhang}, Shuang-Nan and {Santangelo}, Andrea and {Xu}, Yupeng and {Feng}, Hua and {Lu}, Fangjun and {Chen}, Yong and {Ge}, Mingyu and {Nandra}, Kirpal and {Wu}, Xin and {Feroci}, Marco and {Hernanz}, Margarita and {Liu}, Congzhan and {He}, Huilin and {Wang}, Yusa and {Jiang}, Weichun and {Cui}, Weiwei and {Yang}, Yanji and {Wang}, Juan and {Li}, Wei and {Li}, Hong and {Du}, Yuanyuan and {Liu}, Xiaohua and {Meng}, Bin and {Wen}, Xiangyang and {Zhang}, Aimei and {Ma}, Jia and {Li}, Maoshun and {Li}, Gang and {Qi}, Liqiang and {Sun}, Jianchao and {Luo}, Tao and {Liu}, Hongwei and {Liu}, Xiaojing and {Zhang}, Fan and {Luo}, Laidan and {Zhu}, Yuxuan and {Zhao}, Zijian and {Sun}, Liang and {Yang}, Xiongtao and {Wu}, Qiong and {Jiang}, Jiechen and {Shi}, Haoli and {Liu}, Jiangtao and {Xu}, Yanbing and {Yang}, Sheng and {Zhang}, Laiyu and {Han}, Dawei and {Gao}, Na and {Huo}, Jia and {Zhang}, Ziliang and {Wang}, Hao and {Zhao}, Xiaofan and {Wang}, Shuo and {Li}, Zhenjie and {Bao}, Ziyu and {Liu}, Yaoguang and {Wang}, Ke and {Wang}, Na and {Wang}, Bo and {Wang}, Langping and {Wang}, Dianlong and {Ding}, Fei and {Sheng}, Lizhi and {Qiang}, Pengfei and {Yan}, Yongqing and {Liu}, Yongan and {Wu}, Zhenyu and {Liu}, Yichen and {Chen}, Hao and {Zhang}, Yacong and {Liu}, Hongbang and {Altmann}, Alexander and {Bechteler}, Thomas and {Burwitz}, Vadim and {Fiorini}, Carlo and {Friedrich}, Peter and {Meidinger}, Norbert and {Strecker}, Rafael and {Baldini}, Luca and {Bellazzini}, Ronaldo and {Bonino}, Raffaella and {Frass{\`a}}, Andrea and {Latronico}, Luca and {Maldera}, Simone and {Manfreda}, Alberto and {Minuti}, Massimo and {Pesce-Rollins}, Melissa and {Sgr{\`o}}, Carmelo and {Tugliani}, Stefano and {Pareschi}, Giovanni and {Basso}, Stefano and {Sironi}, Giorgia and {Spiga}, Daniele and {Tagliaferri}, Gianpiero and {Tykhonov}, Andrii and {Paltani}, St{\`e}phane and {Bozzo}, Enrico and {Tenzer}, Christoph and {Bayer}, J{\"o}rg and {Tuo}, Youli and {Liu}, Honghui and {Zhang}, Yonghe and {Cai}, Zhiming and {Liu}, Huaqiu and {Chen}, Wen and {Wang}, Chunhong and {He}, Tao and {Chen}, Yehai and {Qiu}, Chengbo and {Zhang}, Ye and {Feng}, Jianchao and {Zhu}, Xiaofei and {Zhou}, Heng and {Zheng}, Shijie and {Song}, Liming and {Wang}, Jinzhou and {Jia}, Shumei and {Jiang}, Zewen and {Li}, Xiaobo and {Zhao}, Haisheng and {Guan}, Ju and {Zhang}, Juan and {Li}, Chengkui and {Huang}, Yue and {Liao}, Jinyuan and {You}, Yuan and {Zhang}, Hongmei and {Wang}, Wenshuai and {Wang}, Shuang and {Ou}, Ge and {Hu}, Hao and {Shi}, Jingyan and {Cui}, Tao and {Jiang}, Xiaowei and {Cheng}, Yaodong and {Li}, Haibo and {Xu}, Yanjun and {Zane}, Silvia and {Bambi}, Cosimo and {Bu}, Qingcui and {Dall'Osso}, Simone and {Rosa}, Alessandra De and {Gou}, Lijun and {Guillot}, Sebastien and {Ji}, Long and {Li}, Ang and {Mao}, Jirong and {Patruno}, Alessandro and {Stratta}, Giulia and {Taverna}, Roberto and {Tsygankov}, Sergey and {Uttley}, Phil and {Watts}, Anna L. and {Wu}, Xuefeng and {Xu}, Renxin and {Yi}, Shuxu and {Zhang}, Guobao and {Zhang}, Liang and {Zhao}, Wen and {Zhou}, Ping},
        title = "{The enhanced X-ray Timing and Polarimetry mission{\textemdash}eXTP for launch in 2030}",
      journal = {Science China Physics, Mechanics, and Astronomy},
         year = 2025,
        month = sep,
       volume = {68},
       number = {11},
          eid = {119502},
        pages = {119502},
          doi = {10.1007/s11433-025-2786-6},
archivePrefix = {arXiv},
       eprint = {2506.08101},
 primaryClass = {astro-ph.HE},
       adsurl = {https://ui.adsabs.harvard.edu/abs/2025SCPMA..6819502Z}
}

@inproceedings{kerr2024upgrading,
  title={Upgrading the Gamma-ray Pulsar Timing Array: Data Release 2},
  author={Kerr, M and Parthasarathy, A and Cromartie, T and Fermi-LAT Collaboration and others},
  booktitle={38th International Cosmic Ray Conference},
  pages={1595},
  year={2024}
}

@article{fermi2022gamma,
  title={A gamma-ray pulsar timing array constrains the nanohertz gravitational wave background},
  author={{Fermi-LAT Collaboration*†}},
  journal={Science},
  volume={376},
  number={6592},
  pages={521--523},
  year={2022},
  publisher={American Association for the Advancement of Science}
}

@article{alam2020nanograv,
  title={The NANOGrav 12.5 yr data set: wideband timing of 47 millisecond pulsars},
  author={Alam, Md F and Arzoumanian, Zaven and Baker, Paul T and Blumer, Harsha and Bohler, Keith E and Brazier, Adam and Brook, Paul R and Burke-Spolaor, Sarah and Caballero, Keeisi and Camuccio, Richard S and others},
  journal={The Astrophysical Journal Supplement Series},
  volume={252},
  number={1},
  pages={5},
  year={2020},
  publisher={IOP Publishing}
}

@article{goncharov2021identifying,
  title={Identifying and mitigating noise sources in precision pulsar timing data sets},
  author={Goncharov, Boris and Reardon, DJ and Shannon, RM and Zhu, Xing-Jiang and Thrane, Eric and Bailes, M and Bhat, NDR and Dai, S and Hobbs, G and Kerr, M and others},
  journal={Monthly Notices of the Royal Astronomical Society},
  volume={502},
  number={1},
  pages={478--493},
  year={2021},
  publisher={Oxford University Press}
}

@article{cordes2016frequency,
  title={Frequency-dependent dispersion measures and implications for pulsar timing},
  author={Cordes, James M and Shannon, Ryan M and Stinebring, Daniel R},
  journal={The Astrophysical Journal},
  volume={817},
  number={1},
  pages={16},
  year={2016},
  publisher={IOP Publishing}
}

@article{donner2020dispersion,
  title={Dispersion measure variability for 36 millisecond pulsars at 150 MHz with LOFAR},
  author={Donner, JY and Verbiest, Joris PW and Tiburzi, Caterina and Os{\l}owski, Stefan and K{\"u}nsem{\"o}ller, J{\"o}rn and Nielsen, A-S Bak and Grie{\ss}meier, J-M and Serylak, M and Kramer, M and Anderson, James M and others},
  journal={Astronomy \& Astrophysics},
  volume={644},
  pages={A153},
  year={2020},
  publisher={EDP Sciences}
}

@article{ligo2017basic,
  title={The basic physics of the binary black hole merger GW150914},
  author={{LIGO Scientific and Virgo Collaborations} and Abbott, BP and Abbott, R and Abbott, TD and Abernathy, MR and Acernese, F and Ackley, K and Adams, C and Adams, T and Addesso, P and others},
  journal={Annalen der Physik},
  volume={529},
  number={1-2},
  pages={1600209},
  year={2017},
  publisher={Wiley Online Library}
}

@article{sesana2004low,
  title={Low-frequency gravitational radiation from coalescing massive black hole binaries in hierarchical cosmologies},
  author={Sesana, Alberto and Haardt, Francesco and Madau, Piero and Volonteri, Marta},
  journal={The Astrophysical Journal},
  volume={611},
  number={2},
  pages={623},
  year={2004},
  publisher={IOP Publishing}
}

@article{hellings1983upper,
  title={Upper limits on the isotropic gravitational radiation background from pulsar timing analysis},
  author={Hellings, RW and Downs, GS},
  journal={Astrophysical Journal, Part 2-Letters to the Editor, vol. 265, Feb. 15, 1983, p. L39-L42.},
  volume={265},
  pages={L39--L42},
  year={1983}
}

@misc{enterprise,
  author       = {Justin A. Ellis and Michele Vallisneri and Stephen R. Taylor and Paul T. Baker},
  title        = {ENTERPRISE: Enhanced Numerical Toolbox Enabling a Robust PulsaR Inference SuitE},
  month        = sep,
  year         = 2020,
  howpublished = {Zenodo},
  doi          = {10.5281/zenodo.4059815},
  url          = {https://doi.org/10.5281/zenodo.4059815}
}

@misc{enterprise_extensions,
  author       = {Stephen R. Taylor and Paul T. Baker and Jeffrey S. Hazboun and Joseph Simon and Sarah J. Vigeland},
  title        = {enterprise\_extensions},
  year         = {2021},
  url          = {https://github.com/nanograv/enterprise_extensions},
  note         = {v2.4.3}
}

@misc{Karnesis:2023ras,
    author = "Nikolaos Karnesis and Michael L. Katz and Natalia Korsakova and Jonathan R. Gair and Nikolaos Stergioulas",
    title = "{Eryn : A multi-purpose sampler for Bayesian inference}",
    eprint = "2303.02164",
    archivePrefix = "arXiv",
    primaryClass = "astro-ph.IM",
    month = "3",
    year = "2023",
    note = "[arXiv:2303.02164]"
}

@software{michael_katz_2023_7705496,
  author       = {Michael Katz and
                  Nikolaos Karnesis and
                  Natalia Korsakova},
  title        = {mikekatz04/Eryn: first full release},
  month        = mar,
  year         = 2023,
  publisher    = {Zenodo},
  version      = {v1.0.0},
  doi          = {10.5281/zenodo.7705496},
  url          = {https://doi.org/10.5281/zenodo.7705496}
}

@ARTICLE{2013PASP..125..306F,
       author = {{Foreman-Mackey}, Daniel and {Hogg}, David W. and {Lang}, Dustin and {Goodman}, Jonathan},
        title = "{emcee: The MCMC Hammer}",
      journal = {Publications of the Astronomical Society of the Pacific},
         year = 2013,
        month = mar,
       volume = {125},
       number = {925},
        pages = {306},
          doi = {10.1086/670067},
archivePrefix = {arXiv},
       eprint = {1202.3665},
 primaryClass = {astro-ph.IM},
       adsurl = {https://ui.adsabs.harvard.edu/abs/2013PASP..125..306F}
}

@article{luo2021pint,
  title={PINT: a modern software package for pulsar timing},
  author={Luo, Jing and Ransom, Scott and Demorest, Paul and Ray, Paul S and Archibald, Anne and Kerr, Matthew and Jennings, Ross J and Bachetti, Matteo and van Haasteren, Rutger and Champagne, Chloe A and others},
  journal={The Astrophysical Journal},
  volume={911},
  number={1},
  pages={45},
  year={2021},
  publisher={IOP Publishing}
}

@article{susobhanan2024pint,
  title={PINT: Maximum-likelihood estimation of pulsar timing noise parameters},
  author={Susobhanan, Abhimanyu and Kaplan, David L and Archibald, Anne M and Luo, Jing and Ray, Paul S and Pennucci, Timothy T and Ransom, Scott M and Agazie, Gabriella and Fiore, William and Larsen, Bjorn and others},
  journal={The Astrophysical Journal},
  volume={971},
  number={2},
  pages={150},
  year={2024},
  publisher={IOP Publishing}
}

@article{hobbs2006tempo2,
  title={TEMPO2, a new pulsar-timing package--I. An overview},
  author={Hobbs, GB and Edwards, RT and Manchester, RN},
  journal={Monthly Notices of the Royal Astronomical Society},
  volume={369},
  number={2},
  pages={655--672},
  year={2006},
  publisher={Blackwell Publishing Ltd Oxford, UK}
}

@article{edwards2006tempo2,
  title={TEMPO2, a new pulsar timing package--II. The timing model and precision estimates},
  author={Edwards, Russell T and Hobbs, GB and Manchester, RN},
  journal={Monthly Notices of the Royal Astronomical Society},
  volume={372},
  number={4},
  pages={1549--1574},
  year={2006},
  publisher={Blackwell Publishing Ltd Oxford, UK}
}

@article{hobbs2009tempo2,
  title={TEMPO2: a new pulsar timing package--III. Gravitational wave simulation},
  author={Hobbs, G and Jenet, F and Lee, KJ and Verbiest, JPW and Yardley, D and Manchester, R and Lommen, A and Coles, W and Edwards, R and Shettigara, C},
  journal={Monthly Notices of the Royal Astronomical Society},
  volume={394},
  number={4},
  pages={1945--1955},
  year={2009},
  publisher={Blackwell Publishing Ltd Oxford, UK}
}

@article{agazie2023nanograv,
  title={The NANOGrav 15 yr data set: evidence for a gravitational-wave background},
  author={Agazie, Gabriella and Anumarlapudi, Akash and Archibald, Anne M and Arzoumanian, Zaven and Baker, Paul T and B{\'e}csy, Bence and Blecha, Laura and Brazier, Adam and Brook, Paul R and Burke-Spolaor, Sarah and others},
  journal={The Astrophysical Journal Letters},
  volume={951},
  number={1},
  pages={L8},
  year={2023},
  publisher={American Astronomical Society}
}

@article{antoniadis2023second,
  title={The second data release from the European Pulsar Timing Array-III. Search for gravitational wave signals},
  author={Antoniadis, John and Arumugam, P and Arumugam, S and Babak, S and Bagchi, M and Nielsen, A-S Bak and Bassa, CG and Bathula, A and Berthereau, A and Bonetti, M and others},
  journal={Astronomy \& Astrophysics},
  volume={678},
  pages={A50},
  year={2023},
  publisher={EDP Sciences}
}

@article{reardon2023search,
  title={Search for an isotropic gravitational-wave background with the Parkes Pulsar Timing Array},
  author={Reardon, Daniel J and Zic, Andrew and Shannon, Ryan M and Hobbs, George B and Bailes, Matthew and Di Marco, Valentina and Kapur, Agastya and Rogers, Axl F and Thrane, Eric and Askew, Jacob and others},
  journal={The Astrophysical Journal Letters},
  volume={951},
  number={1},
  pages={L6},
  year={2023},
  publisher={IOP Publishing}
}

@article{xu2023searching,
  title={Searching for the nano-hertz stochastic gravitational wave background with the Chinese pulsar timing array data release I},
  author={Xu, Heng and Chen, Siyuan and Guo, Yanjun and Jiang, Jinchen and Wang, Bojun and Xu, Jiangwei and Xue, Zihan and Caballero, R Nicolas and Yuan, Jianping and Xu, Yonghua and others},
  journal={Research in Astronomy and Astrophysics},
  volume={23},
  number={7},
  pages={075024},
  year={2023},
  publisher={IOP Publishing}
}

@article{antoniadis2023secondII,
  title={The second data release from the European Pulsar Timing Array-II. Customised pulsar noise models for spatially correlated gravitational waves},
  author={Antoniadis, J and Arumugam, P and Arumugam, S and Babak, S and Bagchi, M and Nielsen, A-S Bak and Bassa, CG and Bathula, A and Berthereau, A and Bonetti, M and others},
  journal={Astronomy \& Astrophysics},
  volume={678},
  pages={A49},
  year={2023},
  publisher={EDP Sciences}
}

@article{caballero2016noise,
  title={The noise properties of 42 millisecond pulsars from the European Pulsar Timing Array and their impact on gravitational-wave searches},
  author={Caballero, RN and Lee, KJ and Lentati, L and Desvignes, Gr{\'e}gory and Champion, DJ and Verbiest, JPW and Janssen, GH and Stappers, BW and Kramer, M and Lazarus, P and others},
  journal={Monthly Notices of the Royal Astronomical Society},
  volume={457},
  number={4},
  pages={4421--4440},
  year={2016},
  publisher={The Royal Astronomical Society}
}

@article{hazboun2022detection,
  title={A Detection of Red Noise in PSR J1824--2452A and Projections for PSR B1937+ 21 using NICER X-ray Timing Data},
  author={Hazboun, Jeffrey S and Crump, Jack and Lommen, Andrea N and Montano, Sergio and Berry, Samantha JH and Zeldes, Jesse and Teng, Elizabeth and Ray, Paul S and Kerr, Matthew and Arzoumanian, Zaven and others},
  journal={The Astrophysical Journal},
  volume={928},
  number={1},
  pages={67},
  year={2022},
  publisher={IOP Publishing}
}

@article{zhao2011constraint,
  title={Constraint on the early Universe by relic gravitational waves: From pulsar timing observations},
  author={Zhao, Wen},
  journal={Physical Review D—Particles, Fields, Gravitation, and Cosmology},
  volume={83},
  number={10},
  pages={104021},
  year={2011},
  publisher={APS}
}

@article{damour2005gravitational,
  title={Gravitational radiation from cosmic (super) strings: Bursts, stochastic background, and observational windows},
  author={Damour, Thibault and Vilenkin, Alexander},
  journal={Physical Review D—Particles, Fields, Gravitation, and Cosmology},
  volume={71},
  number={6},
  pages={063510},
  year={2005},
  publisher={APS}
}

@article{arzoumanian2021searching,
  title={Searching for gravitational waves from cosmological phase transitions with the NANOGrav 12.5-year dataset},
  author={Arzoumanian, Zaven and Baker, Paul T and Blumer, Harsha and Becsy, Bence and Brazier, Adam and Brook, Paul R and Burke-Spolaor, Sarah and Charisi, Maria and Chatterjee, Shami and Chen, Siyuan and others},
  journal={Physical review letters},
  volume={127},
  number={25},
  pages={251302},
  year={2021},
  publisher={APS}
}

@article{khmelnitsky2014pulsar,
  title={Pulsar timing signal from ultralight scalar dark matter},
  author={Khmelnitsky, Andrei and Rubakov, Valery},
  journal={Journal of Cosmology and Astroparticle Physics},
  volume={2014},
  number={02},
  pages={019},
  year={2014},
  publisher={IOP Publishing}
}

@article{jenet2006upper,
  title={Upper bounds on the low-frequency stochastic gravitational wave background from pulsar timing observations: Current limits and future prospects},
  author={Jenet, Frederick A and Hobbs, George B and van Straten, Willem and Manchester, Richard N and Bailes, Matthew and Verbiest, JPW and Edwards, Russell T and Hotan, Aidan W and Sarkissian, John M and Ord, Stephen M},
  journal={The Astrophysical Journal},
  volume={653},
  number={2},
  pages={1571},
  year={2006},
  publisher={IOP Publishing}
}

@article{shannon2013gravitational,
  title={Gravitational-wave limits from pulsar timing constrain supermassive black hole evolution},
  author={Shannon, Ryan M and Ravi, Vikram and Coles, WA and Hobbs, George and Keith, MJ and Manchester, RN and Wyithe, J Stuart B and Bailes, M and Bhat, NDR and Burke-Spolaor, Sarah and others},
  journal={Science},
  volume={342},
  number={6156},
  pages={334--337},
  year={2013},
  publisher={American Association for the Advancement of Science}
}

@article{shannon2015gravitational,
  title={Gravitational waves from binary supermassive black holes missing in pulsar observations},
  author={Shannon, Ryan M and Ravi, Vikram and Lentati, LT and Lasky, Paul Daniel and Hobbs, G and Kerr, Matthew and Manchester, Richard Norman and Coles, William A and Levin, Yuri and Bailes, Matthew and others},
  journal={Science},
  volume={349},
  number={6255},
  pages={1522--1525},
  year={2015},
  publisher={American Association for the Advancement of Science}
}

@article{goncharov2021evidence,
  title={On the evidence for a common-spectrum process in the search for the nanohertz gravitational-wave background with the Parkes Pulsar Timing Array},
  author={Goncharov, Boris and Shannon, RM and Reardon, DJ and Hobbs, G and Zic, A and Bailes, M and Cury{\l}o, M and Dai, S and Kerr, M and Lower, ME and others},
  journal={The Astrophysical Journal Letters},
  volume={917},
  number={2},
  pages={L19},
  year={2021},
  publisher={IOP Publishing}
}

@article{van2011placing,
  title={Placing limits on the stochastic gravitational-wave background using European Pulsar Timing Array data},
  author={van Haasteren, Rutger and Levin, Yuri and Janssen, GH and Lazaridis, K and Kramer, Michael and Stappers, BW and Desvignes, G and Purver, MB and Lyne, AG and Ferdman, RD and others},
  journal={Monthly Notices of the Royal Astronomical Society},
  volume={414},
  number={4},
  pages={3117--3128},
  year={2011},
  publisher={The Royal Astronomical Society}
}

@article{lentati2015european,
  title={European pulsar timing array limits on an isotropic stochastic gravitational-wave background},
  author={Lentati, Lindley and Taylor, Stephen R and Mingarelli, Chiara MF and Sesana, Alberto and Sanidas, Sotiris A and Vecchio, Alberto and Caballero, R Nicolas and Lee, KJ and Van Haasteren, Rutger and Babak, Stanislav and others},
  journal={Monthly Notices of the Royal Astronomical Society},
  volume={453},
  number={3},
  pages={2576--2598},
  year={2015},
  publisher={Oxford University Press}
}

@article{chen2021common,
  title={Common-red-signal analysis with 24-yr high-precision timing of the European Pulsar Timing Array: inferences in the stochastic gravitational-wave background search},
  author={Chen, Shiyang and Caballero, RN and Guo, YJ and Chalumeau, A and Liu, K and Shaifullah, G and Lee, KJ and Babak, S and Desvignes, G and Parthasarathy, A and others},
  journal={Monthly Notices of the Royal Astronomical Society},
  volume={508},
  number={4},
  pages={4970--4993},
  year={2021},
  publisher={Oxford University Press}
}

@article{demorest2012limits,
  title={Limits on the stochastic gravitational wave background from the North American nanohertz observatory for gravitational waves},
  author={Demorest, Paul B and Ferdman, Robert D and Gonzalez, ME and Nice, D and Ransom, S and Stairs, IH and Arzoumanian, Z and Brazier, A and Burke-Spolaor, S and Chamberlin, SJ and others},
  journal={The Astrophysical Journal},
  volume={762},
  number={2},
  pages={94},
  year={2012},
  publisher={IOP Publishing}
}

@article{arzoumanian2018nanograv,
  title={The NANOGrav 11 year data set: pulsar-timing constraints on the stochastic gravitational-wave background},
  author={Arzoumanian, Zaven and Baker, Paul T and Brazier, Adam and Burke-Spolaor, Sarah and Chamberlin, Sydney J and Chatterjee, Shami and Christy, Brian and Cordes, James M and Cornish, Neil J and Crawford, Fronefield and others},
  journal={The Astrophysical Journal},
  volume={859},
  number={1},
  pages={47},
  year={2018},
  publisher={IOP Publishing}
}

@article{arzoumanian2020nanograv,
  title={The NANOGrav 12.5 yr data set: search for an isotropic stochastic gravitational-wave background},
  author={Arzoumanian, Zaven and Baker, Paul T and Blumer, Harsha and B{\'e}csy, Bence and Brazier, Adam and Brook, Paul R and Burke-Spolaor, Sarah and Chatterjee, Shami and Chen, Siyuan and Cordes, James M and others},
  journal={The Astrophysical journal letters},
  volume={905},
  number={2},
  pages={L34},
  year={2020},
  publisher={IOP Publishing}
}

@article{verbiest2016international,
  title={The international pulsar timing array: first data release},
  author={Verbiest, JPW and Lentati, L and Hobbs, G and van Haasteren, Rutger and Demorest, Paul B and Janssen, GH and Wang, J-B and Desvignes, G and Caballero, RN and Keith, MJ and others},
  journal={Monthly Notices of the Royal Astronomical Society},
  volume={458},
  number={2},
  pages={1267--1288},
  year={2016},
  publisher={The Royal Astronomical Society}
}

@article{antoniadis2022international,
  title={The International Pulsar Timing Array second data release: Search for an isotropic gravitational wave background},
  author={Antoniadis, John and Arzoumanian, Z and Babak, S and Bailes, M and Bak Nielsen, AS and Baker, P\_T and Bassa, C\_G and B{\'e}csy, B and Berthereau, A and Bonetti, M and others},
  journal={Monthly Notices of the Royal Astronomical Society},
  volume={510},
  number={4},
  pages={4873--4887},
  year={2022},
  publisher={Oxford University Press}
}

@article{siemens2013stochastic,
  title={The stochastic background: scaling laws and time to detection for pulsar timing arrays},
  author={Siemens, Xavier and Ellis, Justin and Jenet, Fredrick and Romano, Joseph D},
  journal={Classical and Quantum Gravity},
  volume={30},
  number={22},
  pages={224015},
  year={2013},
  publisher={IOP Publishing}
}

@article{pol2021astrophysics,
  title={Astrophysics milestones for pulsar timing array gravitational-wave detection},
  author={Pol, Nihan S and Taylor, Stephen R and Kelley, Luke Zoltan and Vigeland, Sarah J and Simon, Joseph and Chen, Siyuan and Arzoumanian, Zaven and Baker, Paul T and B{\'e}csy, Bence and Brazier, Adam and others},
  journal={The Astrophysical Journal Letters},
  volume={911},
  number={2},
  pages={L34},
  year={2021},
  publisher={IOP Publishing}
}

@article{jenet2004constraining,
  title={Constraining the properties of supermassive black hole systems using pulsar timing: application to 3C 66B},
  author={Jenet, Fredrick A and Lommen, Andrea and Larson, Shane L and Wen, Linqing},
  journal={The Astrophysical Journal},
  volume={606},
  number={2},
  pages={799},
  year={2004},
  publisher={IOP Publishing}
}

@article{white1978core,
  title={Core condensation in heavy halos: a two-stage theory for galaxy formation and clustering},
  author={White, Simon DM and Rees, Martin J},
  journal={Monthly Notices of the Royal Astronomical Society},
  volume={183},
  number={3},
  pages={341--358},
  year={1978},
  publisher={Oxford University Press Oxford, UK}
}

@article{begelman1980massive,
  title={Massive black hole binaries in active galactic nuclei},
  author={Begelman, Mitchell C and Blandford, Roger D and Rees, Martin J},
  journal={Nature},
  volume={287},
  number={5780},
  pages={307--309},
  year={1980},
  publisher={Nature Publishing Group UK London}
}

@article{jaffe2003gravitational,
  title={Gravitational waves probe the coalescence rate of massive black hole binaries},
  author={Jaffe, Andrew H and Backer, Donald C},
  journal={The Astrophysical Journal},
  volume={583},
  number={2},
  pages={616},
  year={2003},
  publisher={IOP Publishing}
}

@article{sesana2008stochastic,
  title={The stochastic gravitational-wave background from massive black hole binary systems: implications for observations with Pulsar Timing Arrays},
  author={Sesana, Alberto and Vecchio, Alberto and Colacino, Carlo Nicola},
  journal={Monthly Notices of the Royal Astronomical Society},
  volume={390},
  number={1},
  pages={192--209},
  year={2008},
  publisher={Blackwell Publishing Ltd Oxford, UK}
}

@article{sesana2013insights,
  title={Insights into the astrophysics of supermassive black hole binaries from pulsar timing observations},
  author={Sesana, A},
  journal={Classical and Quantum Gravity},
  volume={30},
  number={22},
  pages={224014},
  year={2013},
  publisher={IOP Publishing}
}

@article{rajagopal1994ultra,
  title={Ultra-low frequency gravitational radiation from massive black hole binaries},
  author={Rajagopal, Mohan and Romani, Roger W},
  journal={arXiv preprint astro-ph/9412038},
  year={1994}
}

@article{wyithe2003low,
  title={Low-frequency gravitational waves from massive black hole binaries: predictions for LISA and pulsar timing arrays},
  author={Wyithe, J Stuart B and Loeb, Abraham},
  journal={The Astrophysical Journal},
  volume={590},
  number={2},
  pages={691},
  year={2003},
  publisher={IOP Publishing}
}

@article{ravi2015prospects,
  title={Prospects for gravitational-wave detection and supermassive black hole astrophysics with pulsar timing arrays},
  author={Ravi, V and Wyithe, JSB and Shannon, RM and Hobbs, G},
  journal={Monthly Notices of the Royal Astronomical Society},
  volume={447},
  number={3},
  pages={2772--2783},
  year={2015},
  publisher={Oxford University Press}
}

@article{burke2019astrophysics,
  title={The astrophysics of nanohertz gravitational waves},
  author={Burke-Spolaor, Sarah and Taylor, Stephen R and Charisi, Maria and Dolch, Timothy and Hazboun, Jeffrey S and Holgado, A Miguel and Kelley, Luke Zoltan and Lazio, T Joseph W and Madison, Dustin R and McMann, Natasha and others},
  journal={The Astronomy and astrophysics review},
  volume={27},
  number={1},
  pages={5},
  year={2019},
  publisher={Springer}
}

@article{kibble1976topology,
  title={Topology of cosmic domains and strings},
  author={Kibble, Thomas WB},
  journal={Journal of Physics A: Mathematical and General},
  volume={9},
  number={8},
  pages={1387},
  year={1976},
  publisher={IOP Publishing}
}

@article{damour2000gravitational,
  title={Gravitational wave bursts from cosmic strings},
  author={Damour, Thibault and Vilenkin, Alexander},
  journal={Physical Review Letters},
  volume={85},
  number={18},
  pages={3761},
  year={2000},
  publisher={APS}
}

@article{siemens2007gravitational,
  title={Gravitational-wave stochastic background from cosmic strings},
  author={Siemens, Xavier and Mandic, Vuk and Creighton, Jolien},
  journal={Physical Review Letters},
  volume={98},
  number={11},
  pages={111101},
  year={2007},
  publisher={APS}
}

@article{olmez2010gravitational,
  title={Gravitational-wave stochastic background from kinks and cusps on cosmic strings},
  author={{\"O}lmez, S and Mandic, Vuk and Siemens, X},
  journal={Physical Review D—Particles, Fields, Gravitation, and Cosmology},
  volume={81},
  number={10},
  pages={104028},
  year={2010},
  publisher={APS}
}

@article{sanidas2012constraints,
  title={Constraints on cosmic string tension imposed by the limit on the stochastic gravitational wave background from the European Pulsar Timing Array},
  author={Sanidas, Sotirios A and Battye, Richard A and Stappers, Benjamin W},
  journal={Physical Review D—Particles, Fields, Gravitation, and Cosmology},
  volume={85},
  number={12},
  pages={122003},
  year={2012},
  publisher={APS}
}

@article{caprini2010detection,
  title={Detection of gravitational waves from the QCD phase transition with pulsar timing arrays},
  author={Caprini, Chiara and Durrer, Ruth and Siemens, Xavier},
  journal={Physical Review D—Particles, Fields, Gravitation, and Cosmology},
  volume={82},
  number={6},
  pages={063511},
  year={2010},
  publisher={APS}
}

@article{xue2021constraining,
  title={Constraining cosmological phase transitions with the parkes pulsar timing array},
  author={Xue, Xiao and Bian, Ligong and Shu, Jing and Yuan, Qiang and Zhu, Xingjiang and Bhat, ND Ramesh and Dai, Shi and Feng, Yi and Goncharov, Boris and Hobbs, George and others},
  journal={Physical Review Letters},
  volume={127},
  number={25},
  pages={251303},
  year={2021},
  publisher={APS}
}

@article{grishchuk2005relic,
  title={Relic gravitational waves and cosmology},
  author={Grishchuk, Leonid P},
  journal={Physics-Uspekhi},
  volume={48},
  number={12},
  pages={1235},
  year={2005},
  publisher={IOP Publishing}
}

@article{zhao2013constraints,
  title={Constraints of relic gravitational waves by pulsar timing arrays:<? format?> Forecasts for the FAST and SKA projects},
  author={Zhao, Wen and Zhang, Yang and You, Xiao-Peng and Zhu, Zong-Hong},
  journal={Physical Review D—Particles, Fields, Gravitation, and Cosmology},
  volume={87},
  number={12},
  pages={124012},
  year={2013},
  publisher={APS}
}

@article{lasky2016gravitational,
  title={Gravitational-wave cosmology across 29 decades in frequency},
  author={Lasky, Paul D and Mingarelli, Chiara MF and Smith, Tristan L and Giblin Jr, John T and Thrane, Eric and Reardon, Daniel J and Caldwell, Robert and Bailes, Matthew and Bhat, ND Ramesh and Burke-Spolaor, Sarah and others},
  journal={Physical Review X},
  volume={6},
  number={1},
  pages={011035},
  year={2016},
  publisher={APS}
}

@article{galtier2017turbulence,
  title={Turbulence of weak gravitational waves in the early universe},
  author={Galtier, S{\'e}bastien and Nazarenko, Sergey V},
  journal={Physical review letters},
  volume={119},
  number={22},
  pages={221101},
  year={2017},
  publisher={APS}
}

@article{gold1969rotating,
  title={Rotating neutron stars and the nature of pulsars},
  author={Gold, Thomas},
  journal={Nature},
  volume={221},
  number={5175},
  pages={25--27},
  year={1969},
  publisher={Nature Publishing Group UK London}
}

@article{jenet2005detecting,
  title={Detecting the stochastic gravitational wave background using pulsar timing},
  author={Jenet, Fredrick A and Hobbs, George B and Lee, KJ and Manchester, Richard N},
  journal={The Astrophysical Journal},
  volume={625},
  number={2},
  pages={L123},
  year={2005},
  publisher={IOP Publishing}
}

@inproceedings{mushotzky2018axis,
  title={AXIS: a probe class next generation high angular resolution x-ray imaging satellite},
  author={Mushotzky, R},
  booktitle={Space Telescopes and Instrumentation 2018: Ultraviolet to Gamma Ray},
  volume={10699},
  pages={570--591},
  year={2018},
  organization={SPIE}
}

@article{nandra2013hot,
  title={The Hot and Energetic Universe: A White Paper presenting the science theme motivating the Athena+ mission},
  author={Nandra, Kirpal and Barret, Didier and Barcons, Xavier and Fabian, Andy and Herder, Jan-Willem den and Piro, Luigi and Watson, Mike and Adami, Christophe and Aird, James and Afonso, Jose Manuel and others},
  journal={arXiv preprint arXiv:1306.2307},
  year={2013}
}

@article{zhang2019enhanced,
  title={The enhanced X-ray Timing and Polarimetry mission—eXTP},
  author={Zhang, ShuangNan and Santangelo, Andrea and Feroci, Marco and Xu, YuPeng and Lu, FangJun and Chen, Yong and Feng, Hua and Zhang, Shu and Brandt, S{\o}ren and Hernanz, Margarita and others},
  journal={Science China Physics, Mechanics \& Astronomy},
  volume={62},
  pages={1--25},
  year={2019},
  publisher={Springer}
}

@ARTICLE{1994ApJ...428..713K,
       author = {{Kaspi}, V.~M. and {Taylor}, J.~H. and {Ryba}, M.~F.},
        title = "{High-Precision Timing of Millisecond Pulsars. III. Long-Term Monitoring of PSRs B1855+09 and B1937+21}",
      journal = {The Astrophysical Journal},
         year = 1994,
        month = jun,
       volume = {428},
        pages = {713},
          doi = {10.1086/174280},
       adsurl = {https://ui.adsabs.harvard.edu/abs/1994ApJ...428..713K}
}

@ARTICLE{2009MNRAS.400..951V,
       author = {{Verbiest}, J.~P.~W. and {Bailes}, M. and {Coles}, W.~A. and {Hobbs}, G.~B. and {van Straten}, W. and {Champion}, D.~J. and {Jenet}, F.~A. and {Manchester}, R.~N. and {Bhat}, N.~D.~R. and {Sarkissian}, J.~M. and {Yardley}, D. and {Burke-Spolaor}, S. and {Hotan}, A.~W. and {You}, X.~P.},
        title = "{Timing stability of millisecond pulsars and prospects for gravitational-wave detection}",
      journal = {Monthly Notices of the Royal Astronomical Society},
         year = 2009,
        month = dec,
       volume = {400},
       number = {2},
        pages = {951-968},
          doi = {10.1111/j.1365-2966.2009.15508.x},
archivePrefix = {arXiv},
       eprint = {0908.0244},
 primaryClass = {astro-ph.GA},
       adsurl = {https://ui.adsabs.harvard.edu/abs/2009MNRAS.400..951V}
}

@ARTICLE{2024Univ...10..174Z,
       author = {{Zheng}, Shijie and {Han}, Dawei and {Xu}, Heng and {Lee}, Kejia and {Yuan}, Jianping and {Wang}, Haoxi and {Ge}, Mingyu and {Zhang}, Liang and {Li}, Yongye and {Yin}, Yitao and {Ma}, Xiang and {Chen}, Yong and {Zhang}, Shuangnan},
        title = "{New Timing Results of MSPs from NICER Observations}",
      journal = {Universe},
         year = 2024,
        month = apr,
       volume = {10},
       number = {4},
          eid = {174},
        pages = {174},
          doi = {10.3390/universe10040174},
archivePrefix = {arXiv},
       eprint = {2404.16263},
 primaryClass = {astro-ph.HE},
       adsurl = {https://ui.adsabs.harvard.edu/abs/2024Univ...10..174Z}
}

@ARTICLE{2019MNRAS.490.4666P,
       author = {{Perera}, B.~B.~P. and {DeCesar}, M.~E. and {Demorest}, P.~B. and {Kerr}, M. and {Lentati}, L. and {Nice}, D.~J. and {Os{\l}owski}, S. and {Ransom}, S.~M. and {Keith}, M.~J. and {Arzoumanian}, Z. and {Bailes}, M. and {Baker}, P.~T. and {Bassa}, C.~G. and {Bhat}, N.~D.~R. and {Brazier}, A. and {Burgay}, M. and {Burke-Spolaor}, S. and {Caballero}, R.~N. and {Champion}, D.~J. and {Chatterjee}, S. and {Chen}, S. and {Cognard}, I. and {Cordes}, J.~M. and {Crowter}, K. and {Dai}, S. and {Desvignes}, G. and {Dolch}, T. and {Ferdman}, R.~D. and {Ferrara}, E.~C. and {Fonseca}, E. and {Goldstein}, J.~M. and {Graikou}, E. and {Guillemot}, L. and {Hazboun}, J.~S. and {Hobbs}, G. and {Hu}, H. and {Islo}, K. and {Janssen}, G.~H. and {Karuppusamy}, R. and {Kramer}, M. and {Lam}, M.~T. and {Lee}, K.~J. and {Liu}, K. and {Luo}, J. and {Lyne}, A.~G. and {Manchester}, R.~N. and {McKee}, J.~W. and {McLaughlin}, M.~A. and {Mingarelli}, C.~M.~F. and {Parthasarathy}, A.~P. and {Pennucci}, T.~T. and {Perrodin}, D. and {Possenti}, A. and {Reardon}, D.~J. and {Russell}, C.~J. and {Sanidas}, S.~A. and {Sesana}, A. and {Shaifullah}, G. and {Shannon}, R.~M. and {Siemens}, X. and {Simon}, J. and {Spiewak}, R. and {Stairs}, I.~H. and {Stappers}, B.~W. and {Swiggum}, J.~K. and {Taylor}, S.~R. and {Theureau}, G. and {Tiburzi}, C. and {Vallisneri}, M. and {Vecchio}, A. and {Wang}, J.~B. and {Zhang}, S.~B. and {Zhang}, L. and {Zhu}, W.~W. and {Zhu}, X.~J.},
        title = "{The International Pulsar Timing Array: second data release}",
      journal = {Monthly Notices of the Royal Astronomical Society},
         year = 2019,
        month = dec,
       volume = {490},
       number = {4},
        pages = {4666-4687},
          doi = {10.1093/mnras/stz2857},
archivePrefix = {arXiv},
       eprint = {1909.04534},
 primaryClass = {astro-ph.HE},
       adsurl = {https://ui.adsabs.harvard.edu/abs/2019MNRAS.490.4666P}
}

@ARTICLE{2006MNRAS.369..655H,
       author = {{Hobbs}, G.~B. and {Edwards}, R.~T. and {Manchester}, R.~N.},
        title = "{TEMPO2, a new pulsar-timing package - I. An overview}",
      journal = {Monthly Notices of the Royal Astronomical Society},
         year = 2006,
        month = jun,
       volume = {369},
       number = {2},
        pages = {655-672},
          doi = {10.1111/j.1365-2966.2006.10302.x},
archivePrefix = {arXiv},
       eprint = {astro-ph/0603381},
 primaryClass = {astro-ph},
       adsurl = {https://ui.adsabs.harvard.edu/abs/2006MNRAS.369..655H}
}

@ARTICLE{2006MNRAS.372.1549E,
       author = {{Edwards}, R.~T. and {Hobbs}, G.~B. and {Manchester}, R.~N.},
        title = "{TEMPO2, a new pulsar timing package - II. The timing model and precision estimates}",
      journal = {Monthly Notices of the Royal Astronomical Society},
         year = 2006,
        month = nov,
       volume = {372},
       number = {4},
        pages = {1549-1574},
          doi = {10.1111/j.1365-2966.2006.10870.x},
archivePrefix = {arXiv},
       eprint = {astro-ph/0607664},
 primaryClass = {astro-ph},
       adsurl = {https://ui.adsabs.harvard.edu/abs/2006MNRAS.372.1549E}
}

@book{jeffreys1998theory,
  title={The theory of probability},
  author={Jeffreys, Harold},
  year={1998},
  publisher={OuP Oxford}
}

@ARTICLE{2025RAA....25c5022C,
       author = {{Caballero}, R. Nicolas and {Xu}, Heng and {Lee}, Kejia and {Chen}, Siyuan and {Guo}, Yanjun and {Jiang}, Jinchen and {Wang}, Bojun and {Xu}, Jiangwei and {Xue}, Zihan},
        title = "{Chinese Pulsar Timing Array Upper Limits on Microhertz Gravitational Waves from Supermassive Black-hole Binaries Using PSR J1713+0747 FAST Data}",
      journal = {Research in Astronomy and Astrophysics},
         year = 2025,
        month = mar,
       volume = {25},
       number = {3},
          eid = {035022},
        pages = {035022},
          doi = {10.1088/1674-4527/adb5d6},
archivePrefix = {arXiv},
       eprint = {2502.09275},
 primaryClass = {astro-ph.HE},
       adsurl = {https://ui.adsabs.harvard.edu/abs/2025RAA....25c5022C}
}

@ARTICLE{2010arXiv1008.1782C,
       author = {{Corbin}, Vincent and {Cornish}, Neil J.},
        title = "{Pulsar Timing Array Observations of Massive Black Hole Binaries}",
      journal = {arXiv e-prints},
         year = 2010,
        month = aug,
          eid = {arXiv:1008.1782},
        pages = {arXiv:1008.1782},
          doi = {10.48550/arXiv.1008.1782},
archivePrefix = {arXiv},
       eprint = {1008.1782},
 primaryClass = {astro-ph.HE},
       adsurl = {https://ui.adsabs.harvard.edu/abs/2010arXiv1008.1782C}
}
\bibliographystyle{aasjournalv7}



\end{document}